\documentclass[twocolumn,showpacs,preprintnumbers,amsmath,amssymb]{revtex4}

\usepackage{graphicx}% Include figure files
\usepackage{dcolumn}% Align table columns on decimal point
\usepackage{bm}% bold math
\usepackage{dsfont}
\begin{document}  

\title{Linking confinement to spectral properties of the Dirac operator}

\author{Christof Gattringer
\vspace{2mm}}

\affiliation{Institut f\"ur Physik, FB Theoretische Physik, 
Universit\"at Graz, A-8010 Graz, Austria}

\begin{abstract}
We represent Polyakov loops and their correlators as spectral sums 
of eigenvalues and
eigenmodes of the lattice Dirac operator. The deconfinement transition 
of pure gauge theory is characterized as a change in the response of 
moments of eigenvalues to varying the boundary conditions of 
the Dirac operator. We 
argue that the potential between 
static quarks is linked to spatial correlations of Dirac eigenvectors. 
\end{abstract}
\pacs{12.38.Aw, 11.15.Ha, 11.10.Wx}
\keywords{Confinement, Polyakov loop, lattice gauge theory}
\maketitle

\vskip4mm
\noindent
{\bf Introductory remarks}
\vskip1mm
\noindent
Confinement and chiral symmetry breaking are two of the central features of 
QCD. At the QCD finite temperature transition chiral symmetry is restored and 
the theory deconfines. Numerical simulations in lattice QCD indicate that 
the critical temperature $T_c$ 
is the same for both transitions. Thus it is widely
believed that there must be a mechanism linking the two phenomena. However,
so far there is no generally accepted picture for such a link. 

For chiral symmetry breaking an important connection between the 
order parameter, the chiral condensate, and spectral properties of the Dirac
operator is known. 
The Banks-Casher formula \cite{baca} links the chiral condensate to the density
of Dirac eigenvalues at the origin. Building on this connection it has been 
suggested \cite{instantons}
that a liquid of objects with topological charge, e.g., instantons, 
could lead to a non-vanishing density of eigenvalues and thus such topological 
objects are candidates for vacuum excitations responsible for chiral symmetry 
breaking. Above $T_c$ the topological
objects are expected to rearrange and a spectral gap opens up. The chiral
condensate vanishes and chiral symmetry is restored.

Concerning confinement so far no signature in spectral properties of the Dirac
operator is known. On the other hand it is obvious that such signatures must
exist. The inverse Dirac operator, i.e., the quark propagator, clearly 
knows about confinement properties. In this letter we present an attempt to
identify spectral signatures of the Dirac operator which are related to
confinement.   

Our starting point are Polyakov loops on a euclidean lattice. 
A Polyakov loop is defined as the ordered product of temporal link 
variables at a fixed spatial position $\vec{x}$,
\begin{equation}
L(\vec{x}) \; = \; 
\mbox{tr}_c \prod_{s=1}^{N} U_4(\vec{x},s) \; , 
\end{equation}
where the $N$ denotes the number of lattice points in time direction
and $\mbox{tr}_c$ is the trace over color indices. For pure gauge theory
the action is invariant under center rotations, while the Polyakov loop
is not. The QCD phase transition can be viewed as spontaneous breaking
of the center symmetry and the Polyakov loop is the corresponding order
parameter \cite{znbreaking}. In the confined phase it vanishes, but above 
$T_c$, where the theory deconfines, the Polyakov loop acquires a non-vanishing
expectation value. When fermions are coupled, the center symmetry is 
broken explicitly by the Dirac operator and the Polyakov loop
cannot be used as order parameter then. In the dynamical case
correlators of Polyakov loops are related to the potential between static 
quark sources and thus linked to confinement.

Working in the lattice regularization, we express 
the Polyakov loop and its correlators
as a spectral sum of eigenvalues and eigenvectors of 
the Dirac operator with different boundary conditions. If the Polyakov 
loop is averaged over space, our formula contains only moments of  
eigenvalues. The deconfining transition of QCD can then be seen as a 
change in the response of these moments to changing boundary conditions 
of the Dirac operator. We furthermore show, that the static potential is 
connected to spatial correlations of the eigenvectors. 

\vskip4mm
\noindent
{\bf Dirac operator and Polyakov loops}
\vskip1mm
\noindent
To be specific, we work on a lattice of size $L^3 \times N$ and 
for the gauge field (SU$(n)$, $n$ arbitrary) use boundary conditions which are periodic 
in all four directions. We base our discussion on the Wilson 
Dirac operator 
\begin{equation}
D(x|y) = (4 + m) \delta_{x,y} \; - \; \frac{1}{2} 
\sum_{\mu = \pm 1}^{\pm4} [ 1 \mp \gamma_\mu ] \, U_\mu(x) \, 
\delta_{x + \hat{\mu}, y} \; ,
\label{wilsondirac}
\end{equation}
where we use $U_{-\mu}(x) \equiv U_\mu(x - \hat{\mu})^\dagger$. 
At the moment we consider boundary conditions for the Dirac operator
which are periodic for all four directions. We remark that our
construction goes through unchanged when a chemical potential is coupled in 
the standard way \cite{haka}.

The hopping terms of the Wilson Dirac operator (\ref{wilsondirac}) 
connect nearest neighbors. When powers of $D$ are considered, these terms
combine to chains of hops on the lattice. Along these chains 
products of the link variables $U_\mu(x)$ and of the matrices 
$[1 \mp \gamma_\mu]/2$ are collected.
Taking the $m$-th power will give rise to chains with a maximal length 
of $m$ steps (in chains shorter than $m$ the on-site term appears as a
factor). We now consider the $N$-th power of $D$, where $N$ is the 
temporal extent of our lattice. Thus we will encounter chains with a
maximum length of $N$. Furthermore we set the two space-time arguments
of $D$ to the same value, $y = x$, such that we pick up only closed chains, 
i.e., loops starting and ending at $x$. Among these are the loops where 
only hops in time direction occur such that they close around compact
time. We obtain (the $^*$ denotes complex conjugation)
\begin{eqnarray}
&& \mbox{tr}_d \, \mbox{tr}_c \big[ D^{N}(\vec{x},t|\vec{x},t) \big] \; = \;
\mbox{tr}_d \left[ \frac{1-\gamma_4}{2} \right]^{N}  
\mbox{tr}_c \prod_{s=1}^{N} U_4(\vec{x},s) 
\nonumber
\\
&& + \; 
\mbox{tr}_d \left[ \frac{1+\gamma_4}{2} \right]^{N}  
\mbox{tr}_c \,\prod_{s=0}^{N-1} U_4(\vec{x},N\!-\!s)^\dagger
\; + \; \mbox{other loops} 
\nonumber
\\
&& = \; 2 L(\vec{x}) \; + \; 2 L(\vec{x})^* \; + \; \mbox{other loops} \; .
\label{dandloops}
\end{eqnarray} 
Here $\mbox{tr}_d$ and $\mbox{tr}_c$ denote the traces over 
Dirac and color indices. Evaluating the Dirac trace is 
straightforward, due to the projector property 
$([1 \mp \gamma_4]/2)^2 = [1 \mp \gamma_4]/2 $. 
We stress that the forward and backward running Polyakov loops
are the only ones that wind non-trivially around the compact time
direction.

We now explore the fact that the Polyakov loops respond differently
to a change of the boundary conditions compared to other, non-winding loops.
We can change the temporal boundary condition of the Dirac operator
by multiplying all temporal link variables at $t = N$ with some phase factor 
$z \in \mathds{C}, |z| = 1$,
\begin{equation}
U_4(\vec{x}, N) \; \rightarrow \;  z \, U_4(\vec{x}, N) \quad
\forall \, \vec{x}.
\label{utrafo}
\end{equation}
We evaluate the left-hand side of (\ref{dandloops}) a second time, 
now using the Dirac operator in the field transformed according to
(\ref{utrafo}), which we denote as $D_z$. We obtain:
\begin{equation}
\mbox{tr}_{d,c} \big[ D_z^{N}(\vec{x},t|\vec{x},t) \big] \, = \, 
z \, 2 L(\vec{x}) \, + \, z^* \, 2 L(\vec{x})^* \, + \; \mbox{other loops} \, .
\end{equation}
Only the two Polyakov loops which wind 
non-trivially are altered when changing 
the boundary condition. All other loops cross the last temporal link,
where we put our boundary condition, equally often in both directions (or not
at all). Since
forward and backward oriented link variables are conjugate to each other 
the phase factors cancel. 
This fact is used to get rid of the trivial loops:
\begin{eqnarray}
&&\hspace*{-5mm}
\frac{1}{4} \Big( \, \mbox{tr}_{d,c} 
\big[ D^{N}(\vec{x},t|\vec{x},t) \big] \; - \; \mbox{tr}_{d,c} 
\big[ D_z^{N}(\vec{x},t|\vec{x},t) \big] \Big) 
\nonumber
\\
&& \hspace*{10mm} = \;
\frac{1-z}{2} \, L(\vec{x}) \; + \; \frac{1-z^*}{2} \, L(\vec{x})^* \; .
\label{loopcancel}
\label{sdef}
\end{eqnarray}
Note that the right-hand side is independent of $t$ and thus the 
left-hand side can be evaluated at arbitrary $t$. We use this freedom and
average over $t$, a step which will be convenient later. It is easy to see
that combining periodic boundary conditions with $z = \pm$ i boundary 
conditions (subscripts $\pm$) gives the Polyakov loop:
\begin{eqnarray}
\hspace*{-5mm}&& \qquad L(\vec{x}) \, = \, 
\frac{1}{8N}  \sum_{t=1}^N \Big(  \, 
2 \, \mbox{tr}_{d,c} 
\big[ D^{N}(\vec{x},t|\vec{x},t) \big] \, -
\label{LandD}
\\
\hspace*{-5mm}&&(1\!+\!\mbox{i}) \, \mbox{tr}_{d,c} 
\big[ D_+^{N}(\vec{x},t|\vec{x},t) \big]  
\, - \, (1\!-\!\mbox{i})  \, \mbox{tr}_{d,c} 
\big[ D_-^{N}(\vec{x},t|\vec{x},t) \big] \Big) \, .
\nonumber
\end{eqnarray}
We remark that the choice $z = \pm$ i is not particularly special and also
combinations of other phases can project out the Polyakov loop. 
If one is only interested in the real part of the Polyakov loop, already
the combination of periodic and anti-periodic ($z = -1$) temporal 
boundary conditions is sufficient as can be seen from Eq.~(\ref{loopcancel}).

\vskip4mm
\noindent
{\bf Spectral representation of Polyakov loops}
\vskip1mm
\noindent
We now express the Dirac operator using the spectral theorem. 
Since the Wilson Dirac operator is not a normal operator it cannot be 
unitarily diagonalized and we have to use left- and right-eigenvectors. 
We denote them by $\vec{v}_L$ and $\vec{v}_R$, with $\vec{v}_L$ being a row 
vector, while $\vec{v}_R$ is a column vector. They obey
$\vec{v}_{L,\lambda} \, D =  \lambda \, \vec{v}_{L,\lambda}$ and 
$D \, \vec{v}_{R, \lambda} =  \lambda \, \vec{v}_{R,\lambda}$.
The spectrum is the same for both eigenvalue problems, however, 
not the eigenvectors. The eigenvectors form a bi-orthonormal set, 
$\vec{v}_{L,i} \cdot \vec{v}_{R,j} = \delta_{ij}$,
and the spectral representation of the Dirac operator and its $N$-th power
respectively, are given by
\begin{equation}
D \, = \, \sum_\lambda \lambda \, \vec{v}_{R,\lambda} 
\otimes \vec{v}_{L,\lambda} \; , \;
D^N \, = \, \sum_\lambda \lambda^N \, \vec{v}_{R,\lambda} 
\otimes \vec{v}_{L,\lambda} \;.
\label{specdecompN}
\end{equation}
We remark that for a Dirac operator which is $\gamma_5$-hermitian (such as the
Wilson operator we use here) the situation is simplified further: 
A $\gamma_5$-hermitian operator obeys $\gamma_5 D = D^\dagger \gamma_5$,
implying that the left-eigenvector $\vec{v}_{L,\lambda}$ with 
eigenvalue $\lambda$ is related to the right-eigenvector with eigenvalue 
$\lambda^*$ (for a $\gamma_5$-hermitian Dirac operator the eigenvalues 
come in complex conjugate pairs) via 
$\vec{v}_{L,\lambda} = \; \vec{v}_{R,\lambda^*}^{\;\;\dagger} \, \gamma_5$,
and, after suitable normalization, one can work with the
right-eigenvectors alone. 

Inserting the spectral sum (\ref{specdecompN}) into 
(\ref{LandD}) we end up with our final expression for the local 
Polyakov loop:
\begin{eqnarray}
&& \qquad \qquad L(\vec{x}) \, = \, \frac{1}{8N} 
\Big(  \; 2 \sum_\lambda \lambda^N \, \rho(\vec{x})_\lambda \, -
\label{Plocal}
\\
&& (1+\mbox{i}) \sum_{\lambda_+} \lambda_+^N \, 
\rho_+ (\vec{x})_{\lambda_+}
\, - \,(1-\mbox{i}) \sum_{\lambda_-} \lambda_-^N \,
\rho_- (\vec{x})_{\lambda_-} \,
\Big) \; .
\nonumber
\end{eqnarray}
The first sum is over the eigenvalues $\lambda$ obtained with 
periodic boundary conditions, while the second and third sums are 
for the cases with phase
factors $\pm$ i. We have defined the densities $\rho$ and $\rho_\pm$  
for these boundary conditions (color and Dirac indices are summed) as:
\begin{eqnarray}
\rho(\vec{x})_\lambda  & = &
\sum_{t=1}^N \vec{v}_{L, \lambda}(\vec{x},t) \cdot
\vec{v}_{R,\lambda}(\vec{x},t) \; , 
\nonumber
\\
\rho_\pm(\vec{x})_{\lambda_\pm} & = &
\sum_{t=1}^N \vec{v}_{L, \lambda_\pm}(\vec{x},t) \cdot
\vec{v}_{R,\lambda_\pm}(\vec{x},t) \; .
\label{rhodef}
\end{eqnarray}
We remark that both, the eigenvalues as well as the densities 
$\rho$, $\rho_z$ are gauge invariant. Thus gauge invariance 
is explicitly manifest in (\ref{Plocal}), (\ref{rhodef}) as it should be, since the 
formula contains no approximations whatsoever. 

When using the Polyakov loop as an order parameter for confinement 
(pure gauge theory), it is usually averaged over the spatial 
sites $\vec{x}$ to improve statistics,
\begin{equation}
P \; \equiv \; \frac{1}{V_3} \sum_{\vec{x}} L(\vec{x}) \; ,
\end{equation}
where $V_3$ denotes the spatial volume. When summing over spatial indices as
well, the densities (\ref{rhodef}) turn into the matrix elements
of left- and right-eigenvectors with $i = j$ and 
thus the summed densities equal 1. We obtain for the averaged Polyakov loop,
\begin{equation}
P = \frac{1}{8V} 
\left(  2 \sum_\lambda \lambda^N 
- (1\!+\!\mbox{i})  \sum_{\lambda_+} \lambda_+^N 
- (1\!-\!\mbox{i})  \sum_{\lambda_-} \lambda_-^N \right)\!,
\label{Paver}
\end{equation}
where $V = L^3\,N$ is the total number of lattice points. The equation
expresses the Polyakov loop through moments of the Dirac eigenvalues 
at different boundary conditions. We remark that the individual sums are real
since the eigenvalues come in complex conjugate pairs. 

Let us add a few remarks on the case when an exactly chiral lattice Dirac
operator is used. A chiral operator obeys the Ginsparg Wilson \cite{giwi} equation, 
$\gamma_5 D + D \gamma_5 = D \gamma_5 D$, which governs chiral
symmetry on the lattice. All the properties of the Dirac operator
we have used so far still hold. There is, however, an important qualitative 
difference when using a Ginsparg Wilson Dirac 
operator. A solution of the Ginsparg Wilson equation is not ultra-local, 
i.e., $D(x|y)$ contains not only nearest neighbor hops, but 
paths of all lengths. The
coefficients for the paths decrease exponentially with their length. 
The existence of paths of all lengths implies that $D^N$ will not only 
give rise to non-trivially winding loops that are straight lines in time 
direction, but also ``dressed'' Polyakov loops taking some detour
in spatial directions. These dressed loops are exponentially 
suppressed with length (as long as one keeps away from the Aoki phase - see
\cite{aoki}). However, also the dressed Polyakov loops
are not invariant under $Z_N$ transformations and thus are equally
well suited order parameters as are the straight Polyakov loops.

\vskip4mm
\noindent
{\bf Discussion of the spectral sums}
\vskip1mm
\noindent
We begin the interpretation of our formulae with
the spectral sum (\ref{Paver}) for the averaged Polyakov loop $P$. 
The right-hand side expresses the Polyakov loop as linear combination of
the $N$-th moments of the eigenvalues of the Dirac operator at different
boundary conditions. When crossing the QCD phase transition the Polyakov loop 
acquires a non-vanishing expectation value (pure gauge theory), and
the interplay of the eigenvalues at different boundary conditions
must change at $T_c$. 

An interesting question is which part of the Dirac spectrum carries most of the signal 
for the Polyakov loop. The infrared part of the spectrum (small
eigenvalues) is known to undergo a pronounced change as one crosses
from the confining to the deconfined phase: A gap opens up in the spectrum, 
the density of eigenvalues near the origin vanishes and, according to the
Banks-Casher formula, chiral symmetry is restored. The size of the gap depends
on the relative phase of the Polyakov loop and the Dirac boundary condition
\cite{gascha} . 

Although the low-lying eigenvalues undergo a dramatic change, 
it is not clear whether they give a sizable contribution to the Polyakov loop: 
Due to the large power $N$, the small eigenvalues are strongly
suppressed relative to the bulk of the spectrum where eigenvalues of 
${\cal O}(1)$ occur (in lattice units). On the other hand, since the 
hopping term changes sign under a staggered sign transformation, for each
eigenvalue $\lambda$ also $8 - \lambda$ is an eigenvalue. These mirror images
undergo the same drastic change at the phase transition, but are not
suppressed. Thus small eigenvalues could contribute indirectly through their
mirror images at the cutoff. A second important aspect is that the density of
eigenvalues increases roughly as $|\lambda|^3$.
Thus the infrared part of the spectrum and its mirror image at $8 - \lambda$  
have only a small weight. How much each part of the spectrum contributes
to the spectral sum (\ref{Paver}) can probably only be decided numerically. 

In a first test we summed up the lowest 100 eigenvalues of 
the chirally improved lattice Dirac operator \cite{ci}
for quenched SU(3) configurations on $20^3 \times 6$ 
lattices. This was done for temperatures
below and above $T_c$ and in both cases summing up the low-lying modes gave
only a small fraction of the true Polyakov loop $P$ as determined
directly from the link variables. It is remarkable, however, that in all cases
we checked for $T > T_c$, the spectral sum (\ref{Paver}) evaluated with even 
only the 50 lowest eigenvalues already gets the phase of the Polyakov loop right
(the $Z_3$ symmetry gives rise to 3 preferred phases of the Polyakov loop 
expectation value). 
Unfortunately very little is known about the ultraviolet part of the
spectrum. Numerical evaluation of the eigenvalue spectrum is usually
restricted to the infrared part or to very small lattices due to the
immense computational cost of a complete diagonalization. Random matrix
theory, which is very successful in describing the infrared part of the
spectrum, is valid only up to the Thouless energy and provides no
insight to the ultraviolet part of the spectrum. 

We now turn to interpreting Eq.~(\ref{Plocal}) which is the spectral 
sum for the local Polyakov loop at $\vec{x}$. In its local form the Polyakov 
loop $L(\vec{x})$ can be used for computing the potential $V(r)$ between static 
sources ($r \equiv |\vec{x} - \vec{y}|$),
\begin{equation}
\big\langle L(\vec{x}) \, L (\vec{y})^*  \big\rangle \propto 
\exp\big( -  N \, V(r) \big)  \sim 
\exp\big( -\sigma N r  \big) \, ,
\label{Lcorr}
\end{equation}
where the last expression is the behavior for large $r$,
with $\sigma$ denoting the string tension. When inserting the spectral sum 
(\ref{Plocal}) in the correlator (\ref{Lcorr}), one sees that the static 
potential is related to correlations of the densities $\rho$ and $\rho_\pm$.
In particular correlations of the form, 
$\langle \rho(\vec{x})_\lambda \rho(\vec{y})_{\lambda^\prime}^* \rangle$,  
$\langle \rho(\vec{x})_\lambda  \rho_\pm(\vec{y})_{\lambda_\pm^\prime}^* \rangle$, 
$\langle \rho_\pm(\vec{x})_{\lambda_\pm} \rho_\pm(\vec{y})_{\lambda_\pm^\prime}^*\rangle$,
build up $\exp(- N \, V(|\vec{x} - \vec{y}|))$. 
In the confining phase the string
tension $\sigma$ is non-vanishing, such that the density correlations
are expected to decay exponentially with $|\vec{x} - \vec{y}|$ (until
string breaking sets in). 
In the deconfined phase one has $\sigma = 0$ and the decay of the 
eigenvector density correlators should show  a power law behavior. 

Finally, let us address another possible application of our result: 
The spectral representation (\ref{Plocal}) of 
$L(\vec{x})$ might be an interesting observable also for individual gauge 
configurations. It provides a filter for analyzing topological infrared 
structures of the gauge field. 
Expressing gluonic observables in terms of spectral 
sums and truncating these sums has been applied before
(for examples see, e.g., \cite{lowevalfilt}).
The technique is sometimes referred to as ``low eigenmode filtering''. It makes
use of the fact that the low-lying modes of the Dirac operator are an 
efficient filter for infrared properties of the gauge field. In a thermalized 
configuration, i.e., one not treated with cooling or smearing, these are hidden under 
UV fluctuations and a filter is needed to observe them. Using the low-lying 
Dirac modes has the advantage over smearing or cooling techniques, that the 
gauge field is not altered. Also low-lying modes of the covariant Laplace 
operator have been implemented as a low pass filter \cite{laplacefiltering}.

The newly developed formula (\ref{Plocal}) for the Polyakov loop might be
particularly useful for analyzing properties of so-called 
Kraan-van Baal solutions \cite{kvb}. 
There an object of topological charge 1 is made from $n$ constituents (for
gauge group SU($n$)) and the local Polyakov loop is expected to show a specific 
pattern at the positions of the constituents. For cooled configurations this
pattern has been confirmed \cite{kvbpolyakov}. The expression (\ref{Plocal}) can be
used to study the local behavior of the Polyakov loop directly for
thermalized configurations using low 
eigenmode filtering techniques. Concerning the dominant contributions,
for the local version $L(\vec{x})$ the situation is 
different from the summed Polyakov loop $P$. In the spectral sum
(\ref{Plocal}) the eigenvalues are multiplied with the densities 
$\rho(\vec{x})_\lambda$ and $\rho(\vec{x})_{\lambda_\pm}$. 
For the low-lying eigenvectors these densities are known to be large for 
positions where topological objects are located, while for larger values of
$\lambda, \lambda_\pm$ the 
densities are dominated by fluctuations. Numerical tests of using
(\ref{Plocal}) as an infrared filter are in preparation.   
     
\vskip4mm
\noindent
{\bf Concluding remarks}
\vskip1mm
\noindent
How chiral symmetry breaking reflects itself in spectral properties of
the Dirac operator is well understood. The chiral condensate is related 
to the density of eigenvalues near the origin and random matrix theory 
can be used to describe the behavior of the low-lying eigenvalues. 

Also confinement should have a signature in spectral properties of the 
Dirac operator, which, however, is only vaguely understood so far. In this 
letter we have shown that the Polyakov loop can be expressed through
moments of the Dirac eigenvalues computed with different boundary 
conditions. At the phase transition of pure gauge theory the Polyakov loop
acquires a non-vanishing expectation value and the dependence of these 
moments on the boundary condition changes. 
We also discuss correlators of local Polyakov loops which are related to 
the potential between static quarks. Based on this relation we suggest that 
in the confining phase correlators of Dirac eigenvector densities decay 
exponentially with spatial separation, while a power law should be seen 
in the deconfined phase.   

Some of our spectral sums might get sizable contributions from large
eigenvalues. Analyzing the ultraviolet part of the Dirac spectrum is a challenge
both for numerical and analytical approaches. The results presented here 
are certainly only an initial step indicating which spectral quantities can be studied.
The ambitious goal of such a study could be a link between the spectral
properties in the infrared, which are connected to chiral symmetry breaking,
and the ultraviolet part probably more relevant for confinement.

\vspace{-7mm}
\begin{acknowledgments}
\vspace{-3mm}
The author thanks Falk Bruckmann, Tom DeGrand, Antonio Garcia-Garcia, Christian Lang, 
Michael M\"uller-Preussker, Kim Splittorff and Pierre van Baal
for valuable discussions and remarks. We thank the
Institute for Nuclear Theory at the University of Washington, where part of
this work was done, for its hospitality and the Department of Energy 
for support during that stay.  
\end{acknowledgments}


\begin{thebibliography}{1234567}

\vspace{-1mm}

\bibitem{baca}
T.~Banks and A.~Casher, Nucl.~Phys.~B169, 103 (1980).

\bibitem{instantons}
T.~Sch\"afer,~E.V.~Shuryak,~Rev.~Mod.~Phys.~70,~323~(1998);
D.~Diakonov, Prog.\ Part.\ Nucl.\ Phys.\ 51, 173 (2003).
%%CITATION = HEP-PH 0212026;%%

\bibitem{znbreaking}
L.~McLerran,~B.~Svetitsky,~Phys.~Rev.~D~24,~450~(1981).
%%CITATION = PHRVA,D24,450;%%

\bibitem{haka}
P.~Hasenfratz, F.~Karsch,
%``Chemical Potential On The Lattice,''
Phys.\ Lett.\ B 125, 308 (1983).
%%CITATION = PHLTA,B125,308;%%

\bibitem{giwi}
P.H.~Ginsparg, K.G.~Wilson, Phys.\ Rev.\ D 25, 2649 (1982).
%%CITATION = PHRVA,D25,2649;%%

\bibitem{aoki}
M.~Golterman, Y.~Shamir,
%``Localization in lattice QCD,''
Phys.\ Rev.\ D 68, 074501 (2003);
M.~Golterman, Y.~Shamir, B.~Svetitsky,
%``Mobility edge in lattice QCD,''
Phys.\ Rev.\ D 71 (2005) 071502;
%%CITATION = HEP-LAT 0407021;%%
B.~Svetitsky, Y.~Shamir, M.~Golterman,
PoS LAT2005, 129 (2006).
%%CITATION = HEP-LAT 0508015;%%


\bibitem{gascha}
C.~Gattringer, S.~Schaefer,
Nucl.\ Phys.\ B 654, 30 (2003);
%%CITATION = HEP-LAT 0212029;%%
C.~Gattringer, P.E.L.~Rakow, A.~Sch\"afer, W.~S\"oldner,
Phys.\ Rev.\ D 66, 054502 (2002).
%%CITATION = HEP-LAT 0202009;%%

\bibitem{ci}
C.~Gattringer, Phys.~Rev.~D 63, 114501 (2001);
%%CITATION = HEP-LAT 0003005;%%
C.~Gattringer, I.~Hip, C.B.~Lang, Nucl.~Phys.~B597, 451 (2001).
%%CITATION = HEP-LAT 0007042;%%

\bibitem{lowevalfilt}
I.~Horvath {\it et al.},
Phys.\ Rev.\ D 66, 034501 (2002);
%%CITATION = HEP-LAT 0201008;%%
Phys.\ Rev.\ D 67, 011501 (2003);
%%CITATION = HEP-LAT 0203027;%%
%
C.~Gattringer,
Phys.\ Rev.\ Lett.\ 88 (2002) 221601;
%%CITATION = HEP-LAT 0202002;%%
%
C.~Aubin {\it et al.}  [MILC Collaboration],
Nucl.\ Phys.\ Proc.\ Suppl.\ 140, 626 (2005);
%%CITATION = HEP-LAT 0410024;%%
%
J.~Gattnar {\it et al.},
Nucl.\ Phys.\ B 716, 105 (2005);
%%CITATION = HEP-LAT 0412032;%%
%
C.~Gattringer {\it et al.},
Nucl.\ Phys.\ B 617, 101 (2001);
%%CITATION = HEP-LAT 0107016;%%
Nucl.\ Phys.\ B 618, 205 (2001);
%%CITATION = HEP-LAT 0105023;%%
F.V.~Gubarev, S.M.~Morozov, M.I.~Polikarpov, V.I.~Zakharov,
hep-lat/0505016.
%%CITATION = HEP-LAT 0505016;%%


\bibitem{laplacefiltering}
J.\ Greensite {\it et al},
Phys.\ Rev.\ D71, 114507 (2005);
%%CITATION = HEP-LAT 0504008;%%
%
F.~Bruckmann, E.-M.~Ilgenfritz,
Phys.\ Rev.\ D 72, 114502 (2005);
%%CITATION = HEP-LAT 0509020;%%
PoS LAT2005, 305 (2005);
%%CITATION = HEP-LAT 0509087;%%
Nucl.\ Phys.\ Proc.\ Suppl.\ 153, 33 (2006).
%%CITATION = HEP-LAT 0511030;%%

\bibitem{kvb}
T.C.~Kraan, P.~van~Baal, 
Phys.\ Lett.\ B 428 (1998) 268;
%%CITATION = HEP-TH 9802049;%% 
B 435 (1998) 389,
%%CITATION = HEP-TH 9806034;%%
Nucl.\ Phys.\ B 533 (1998) 627;
%%CITATION = HEP-TH 9805168;%%
K.~Lee, C.~Lu,
Phys.\ Rev.\ D 58 (1998) 1025011.
%%CITATION = HEP-TH 9802108;%% 


\bibitem{kvbpolyakov}
E.-M.\ Ilgenfritz {\it et al}, 
Phys.\ Rev.\ D 66 (2002) 074503.
%%CITATION = HEP-LAT 0206004;%%



\end{thebibliography}
\end{document}